\documentclass{aa}
\usepackage[separate-uncertainty,separate-uncertainty-units=single]{siunitx}
\usepackage{newtxtext,newtxmath,dsfont}
\usepackage[T1]{fontenc}

\DeclareRobustCommand{\VAN}[3]{#2}
\let\VANthebibliography\thebibliography
\def\thebibliography{\DeclareRobustCommand{\VAN}[3]{##3}\VANthebibliography}

\usepackage{graphicx}	%
\usepackage{hyperref}
\usepackage[noabbrev,capitalize]{cleveref}
\hypersetup{
    colorlinks=true,
    linkcolor=blue,
    filecolor=magenta,
    urlcolor=blue,
    citecolor=blue
}
\usepackage{amsmath}	%
\usepackage{tikz}
\usetikzlibrary{positioning}
\usepackage{mathtools}
\usepackage{multirow}
\usepackage{bigdelim}
\usepackage{stfloats}
\usepackage{physics}
\usepackage[shortlabels]{enumitem}
\setlist[enumerate]{nosep}
\usepackage{booktabs}

\usepackage{xspace}
\newcommand*{\borg}{\textsc{borg}\xspace}

\makeatletter
\renewcommand*\aa@pageof{, page \thepage{} of \pageref*{LastPage}}
\makeatother

\defcitealias{wempeConstrainedCosmologicalSimulations2024}{W24}

\DeclareSIUnit{\MSun}{\ensuremath{M_\odot}}
\DeclareSIUnit{\Msun}{\ensuremath{M_\odot}}
\DeclareSIUnit[quantity-product = ]\percent{\char`\%}

\newcommand{\LCDM}{$\Lambda$CDM\xspace}

\begin{document}

   \title{The effect of environment on the mass assembly history of the Milky Way and M31}

   \author{Ewoud
          Wempe \inst{1}\fnmsep\thanks{\email{ewoudwempe@gmail.com}}
          \and
          Amina Helmi \inst{1}
          \and
          Simon D.M. White \inst{2}
          \and
          Jens Jasche \inst{3}
          \and
          Guilhem Lavaux \inst{4}
          }

   \institute{Kapteyn Astronomical Institute, University of Groningen, P.O Box 800, 9700 AV Groningen, The Netherlands
              \and
                Max-Planck-Institut für Astrophysik, Karl-Schwarzschild-Straße 1, 85748 Garching, Germany
                \and
                        The Oskar Klein Centre, Department of Physics, Stockholm University, Albanova University Center, SE 106 91 Stockholm, Sweden
              \and
                  CNRS \& Sorbonne Université, UMR 7095, Institut d’Astrophysique de Paris, 98 bis boulevard Arago, F-75014 Paris, France
                  }
   \date{Received X; accepted Y}

\abstract
{We study the mass growth histories of the halos of Milky Way and M31 analogues formed in constrained cosmological simulations of
the Local Group. These
simulations constitute a fair and representative set of $\Lambda$CDM realisations conditioned on properties of the main Local Group galaxies, such as their masses, relative separation, dynamics and environment. Comparing with isolated analogues extracted from the TNG dark-matter-only simulations,
we find that while our M31 halos have a comparable mass growth history to their isolated counterparts, our Milky Ways typically form earlier and their growth is suppressed at late times. Mass growth associated to major and minor mergers is also biased early for the Milky Way in comparison to M31, with most accretion occurring 1 - 4 Gyr after the Big Bang, and a relatively quiescent history at later times. 32\% of our Milky Ways experienced a Gaia-Enceladus/Sausage (GES)-like merger, while 13\% host an LMC-like object at the present day, with 5\% having both.  In one case, an SMC- and a Sagittarius-analogue are also present, showing that the most important mergers of the Milky Way in its Local Group environment can be reproduced in \LCDM. We find that the material that makes up the Milky Way and M31 halos at the present day first collapsed onto a plane roughly aligned with the Local Sheet and Supergalactic plane; after $z\sim2$, accretion occurred mostly within this plane, with the tidal effects of the heavier companion, M31, significantly impacting the late growth history of the Milky Way.

}


   \maketitle

\section{Introduction}

In recent years, tremendous progress has been made on constraining the assembly history of the Milky Way (MW) using {\it Gaia} data \citep{gaiacollaborationGaiaDataRelease2018,gaiacollaborationGaiaDataRelease2023} in combination with ground-based spectroscopic surveys such as APOGEE \citep{majewskiApachePointObservatory2017}.
We now know that the Galaxy experienced its last major merger approximately 10 Gyr ago, with a system dubbed Gaia-Enceladus \citep{helmiMergerThatLed2018} and which created a kinematic structure known as the Gaia-Sausage \citep{belokurovCoformationDiscStellar2018}, hence the event is often referred to as GES for short. We also know that since then accretion has been primarily driven by minor mergers \citep{helmiStreamsSubstructuresEarly2020,naiduEvidenceH3Survey2020,deasonGalacticArchaeologyGaia2024}.
Although it is possible to find MW-mass systems in cosmological simulations that have had a similarly quiescent history, they are nonetheless not so common. What also appears to be uncommon in such simulations is the presence of systems like the Large and Small Magellanic Clouds (hereafter LMC and SMC respectively).

For example, \citet{evansHowUnusualMilky2020a} using the EAGLE simulations \citep{schayeEAGLEProjectSimulating2015a}, find that 5\% of MW-mass haloes have a GES-like event and no further merger with an equally massive object after $z = 1$, while 16\% have an LMC presently; only 0.65\% satisfy the GES and LMC constraints simultaneously.
The progenitors of the MW's in this latter category appear to be much less massive than average at early times (before the merger with the GES).
The galaxies in the simulations are typically selected as isolated systems, though the impact of a group environment could be substantial \citep[see e.g.][]{carlesiMassAssemblyHistory2020,santistevanFormationTimesBuilding2020}.
In more recent work by \citet{buchMilkyWayestCosmological2024}, a set of MW-like halos were identified in cosmological dark-matter-only simulations hosting both a GES-like event at early times and an LMC at late times.
Depending on the exact requirements on the properties of these objects, as few as 0.75\% of MW candidates were found, with a small subset being in a group-like environment.

In this paper we study the effect of the environment on the mass assembly history of Milky Way, and in particular of its embedding in the Local Group (where it is not the most massive galaxy) using the constrained simulations presented in \citet[hereafter \citetalias{wempeConstrainedCosmologicalSimulations2024}]{wempeConstrainedCosmologicalSimulations2024}. By using a suite of constrained simulations that match the detailed properties of the Local Group, we would like to understand whether to expect the Milky Way to assemble unusually early and how the environment affects its growth through cosmic time.

\section{Methodology}
\subsection{W24 Local Group simulations}
\label{sec:w24sims}
The constrained Local Group simulations presented in \citetalias{wempeConstrainedCosmologicalSimulations2024} constitute a fair and representative set of \LCDM realisations conditional on properties of the Local Group; specifically, they are constrained to match the masses, positions and 3D velocities of the MW-M31 pair, and are also constrained by the observed velocity field within \SI{\sim4}{Mpc}.
Hence, if we compare these simulations with MW-mass or M31-mass galaxies identified in large cosmological simulations of random volumes, we can quantify the effect of their specific environment on the MW and M31 mass assembly histories.

For this work, we extract a set of 224 Local Group realisations\footnote{This selection was made from a homogeneous subsample of from the chains, after excluding 15 cases (out of 239 total), where either the MW or M31 was part of a galaxy pair (defined as having a satellite with mass ratio $>0.3$).} from the chains\footnote{We have run the chains from \citetalias{wempeConstrainedCosmologicalSimulations2024} twice as long to increase the effective sample size. In \cref{app:autocorrelations} we discuss in more detail the degree to which the realisations may be considered as (quasi)independent.} run in \citetalias{wempeConstrainedCosmologicalSimulations2024}.
For each sample, we take the initial condition field, add random small-scale perturbations at scales below those included in the chain simulations, and generate initial conditions at redshift $z=63$.
We describe the exact procedure for the upsampling of the initial conditions in \cref{app:icupsampling}.
We then integrate these initial conditions to $z=0$ using Gadget-4 \citep{springelSimulatingCosmicStructure2021}.
The numerical parameters used in the simulations are summarised in \cref{tab:simsettings}. Although the resolution of our resimulations is relatively low, it is comparable to that of the Millenium-II simulation by \citet{boylan-kolchinResolvingCosmicStructure2009,boylan-kolchinTheresNoPlace2010}; who demonstrated convergence for MW-mass halo growth histories by comparing to the higher resolution Aquarius level-2 simulation suite from \citet{springelAquariusProjectSubhaloes2008}.

We use the FOF group-finder, SUBFIND-HBT and merger tree implementations bundled with Gadget-4 to obtain the group catalogues, subhalo catalogues and the merger trees respectively.
In this paper, we denote the virial mass as $M_\text{200c}$ and define it as the mass within a sphere of average density $200\rho_c$, where $\rho_c$ is the critical density of the Universe\footnote{We have also considered $M_{100c}$, for which all results in this paper remain qualitatively unchanged. Generally, the effects reported in this paper and associated to the environment are enhanced for $M_\text{100c}$ as this encompasses mass in a more extended region.}. The virial masses for the Milky Way and M31 DM halos in our simulations are
$M_\text{200c, MW} = \SI{1.23(0.20:0.22)e12}{\Msun}$ and $M_\text{200c, M31} =  \SI{2.08(0.40:0.39)e12}{\Msun}$ (mean and 16th/84th percentiles ). As stated earlier, our MW and M31 halos are constructed to match the observational constraints, with
their masses reproduced within observational uncertainties in our posterior simulations with Gadget. The distances between the main haloes are \SI{806(124:133)}{kpc}, they are infalling with radial velocities of \SI{-84(32:31)}{km.s^{-1}}, and have tangential velocities of \SI{27(17:16)}{km.s^{-1}}.
The spread in these quantities, especially the radial velocity and distance, is larger than the scatter in the low-resolution runs presented in \citetalias{wempeConstrainedCosmologicalSimulations2024}. This is caused by the (significant) increase in resolution of the resimulations, as well as the added entropy resulting from the augmenting of initial conditions with small-scale random fluctuations.
\begin{table}
  \caption{The numerical parameters used in the Local Group Gadget-4 simulations presented in this paper.}
  \label{tab:simsettings}
  \begin{center}
    \begin{tabular}[c]{lll}
      \toprule
      & HR region & LR region \\
      \midrule
      Particle mass & \SI{1.88e7}{\Msun} & \SI{9.65e9}{\Msun}\\
      Softening length & \SI{1.5}{kpc/h} & \SI{24}{kpc/h}\\
      Box size & \SI{10}{Mpc} & \SI{40}{Mpc} \\
      \bottomrule
    \end{tabular}
  \end{center}
\end{table}

\subsection{IllustrisTNG MW/M31-mass haloes}
To assess the effect of the LG environment, we compare with a sample of MW/M31-mass haloes extracted from the DM-only variant of TNG-50.
We choose a sample similar to the `halo-based' selection from \citet{pillepichMilkyWayAndromeda2023}.
Specifically, we select the central subhalos of FoF groups with a $M_\text{200c}$ between \SI{6e11}{\Msun} and \SI{3e12}{\Msun}.
The \SI{3e12}{\Msun} limit is higher than the \SI{2e12}{\Msun} limit of \citet{pillepichMilkyWayAndromeda2023}, because we find that our M31s can have masses larger than \SI{2e12}{\Msun}, and this selection covers the range of $M_\text{200c}$ we obtain from the \citetalias{wempeConstrainedCosmologicalSimulations2024} sample described in \cref{sec:w24sims}. 
\sisetup
{
list-exponents = combine-bracket ,
product-exponents = combine-bracket ,
range-exponents = combine-bracket,
range-units=single,
range-phrase=~\text{--}~
}

We end up with 257 objects,  similar to the 220 objects that \citet{pillepichMilkyWayAndromeda2023} find with stricter limits in their `halo-based' selection. We also define two subsets to represent MW-mass and M31-mass samples. For this, we require that the $M_\text{200c}$'s, centred on the position of main subhalo's most bound particle, are within $1\sigma$ of the masses of the MW and M31 analogues in the \citetalias{wempeConstrainedCosmologicalSimulations2024} sample. In practice, this means $M_\text{200c,MW} \in \qtyrange{0.98e12}{1.45e12}{\Msun}$ and $M_\text{200c,M31} \in \qtyrange{1.62e12}{2.52e12}{\Msun}$, which gives us 56 MW-mass haloes and 56 M31-mass haloes. Additionally, we require no subhalo of more than half the mass of the primary halo within \SI{1}{Mpc} to select isolated haloes. This removes 4 MW-mass haloes and 3 M31-mass haloes.

We utilise mostly the Sublink merger trees part of TNG-50's public data release  \citep{nelsonIllustrisTNGSimulationsPublic2019}. Although this algorithm differs from the one used for the merger trees for the constrained Local Group simulations, the only quantity we use is the position of the most bound particle, which should be robust for the MW and M31 main progenitors across different subhalo finders and linkers.
However we found that for \SI{\sim10}{\percent} of objects, these merger trees missed a link in the main progenitor branch. We therefore decided to perform an additional correction step: if the sum of the progenitor masses of a particular subhalo in the tree is less than half of the mass of the descendant (meaning it misses a massive progenitor), we search in the neighbouring (comoving) \SI{200}{kpc} for a subhalo with a similar mass and if found, manually add this to the merger tree. We verified by visual inspection of the snapshots that in all 22 such cases, the manually inserted extra link looked robust and that no wrong subhaloes have been linked.

\section{Results}

\subsection{Mass assembly history}
\label{sec:massgrowthhistory}

\begin{figure}
    \centering
    \includegraphics[width=\linewidth]{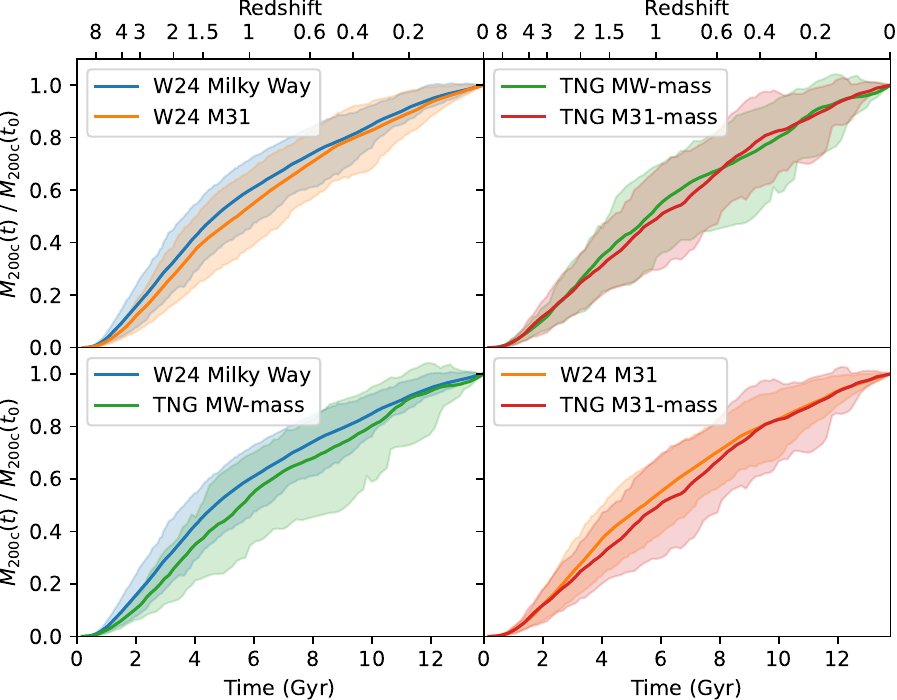}
    \caption{Comparison of the mass growth histories of the Milky Way and M31 galaxies from the constrained simulations of \citetalias{wempeConstrainedCosmologicalSimulations2024}, compared to random Milky Way or M31-mass galaxies from TNG50-DMO. The central lines are the means, and shading indicates the 16-84 percentile regions. The unnormalised mass growth histories are shown in \cref{app:unnormalised}.}
    \label{fig:mencevolution}
\end{figure}

We show the evolution of the virial mass $M_{200c}$ of the MW/M31 halos in \cref{fig:mencevolution}.
In the top left panel, we plotted the mass growth history normalised by present-day mass in the constrained simulations. This figure shows that on average the MW assembles a bit earlier than M31 in our simulations. In \SI{60}{\percent} of cases $a_{50}$ is smaller ($a_X$ is defined as the scale factor at which $X$\% of the present-day mass was in place), and in \SI{69}{\percent} $a_{10}$ is smaller, although the magnitude of the difference is $\lesssim 1\sigma$ of the scatter between realisations.

Some of this bias towards early assembly could in principle be attributed to the Milky Way being less massive (the ratio of the two virial masses is $M_\text{200c,MW}/M_\text{200c,M31} = 0.57\substack{+0.12 \\ -0.16}$ in our suite) since, on average, heavier objects have a later assembly time in \LCDM.
To analyse this in more depth, we consider the mass histories of MW- and M31-mass halos from the TNG-50 DMO-variant, selected as described above.
The top right panel of \cref{fig:mencevolution} compares their mass growth histories; random MW-mass halos indeed assemble slightly earlier on average.
However, the amplitude of the effect is not enough to explain what is seen for our constrained simulations in the top left panel.
Furthermore, comparing MW growth histories between the constrained realisation and TNG samples (bottom left panel), we see that the former are biased towards earlier assembly (although well within the spread), while for M31 the effect of the environment is not apparent, as shown in the bottom right panel.
In summary, it appears that the LG environment causes MW
halos to grow slightly faster at early times and more slowly at late times compared to their isolated counterparts.
It also results in much less scatter in the allowed histories of the MW compared to the TNG sample of MWs. In particular the tail of late-forming MW's which is present in the TNG sample, is absent in our LG sample.

\begin{figure}
    \centering
    \includegraphics[width=\linewidth]{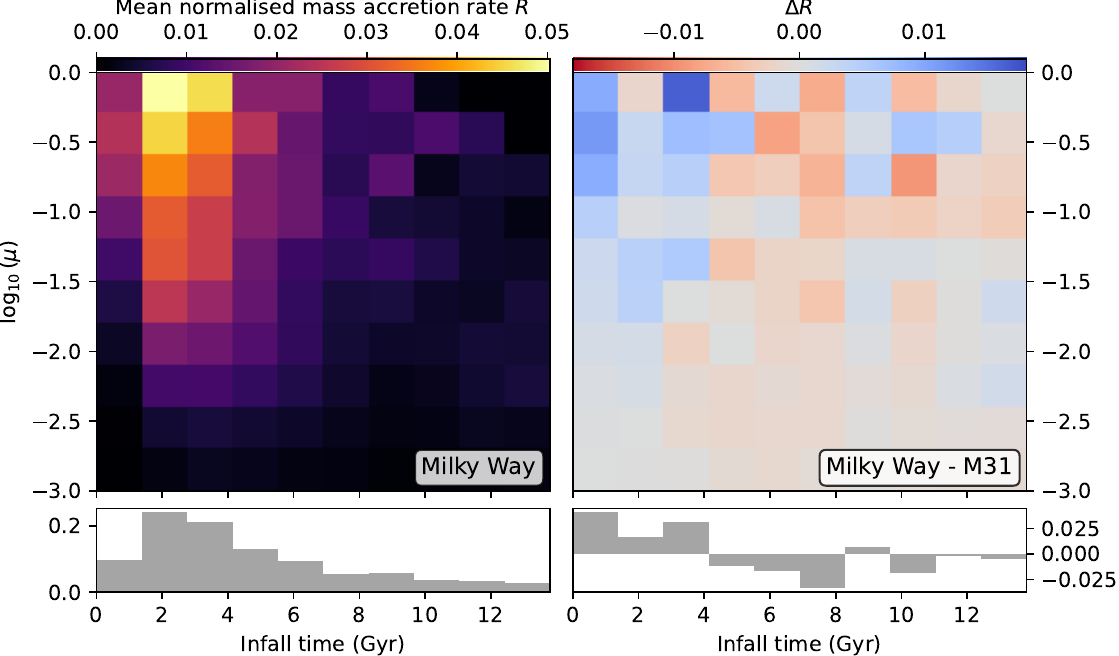}
    \caption{Merger-associated mass accretion rates as a function of infall time and mass ratio ($\mu$). For each bin, the peak masses of the infalling objects are summed and normalised by the total merger-associated mass accretion for the respective halo (MW or M31). This normalised mass accretion rate $R$ is then averaged over simulations, with the resulting values shown by the colours. The left image shows this quantity for MW, and the right image shows the difference with the MW and M31. The histograms at the bottom represent the same quantity, but marginalised over mass ratio, showing the 1D distribution as a function of infall time. }
    \label{fig:massgrowthmassratiosplit}
\end{figure}

The mass growth history is at least in part driven by mergers and accretion events.
In \cref{fig:massgrowthmassratiosplit}, we show the average mass accretion rates via mergers for our MWs as a function of mass ratio and infall time, as well as the difference between the our MWs and M31s.
The mass ratio is defined as the ratio of subhalo to most massive progenitor mass when the subhalo reaches its maximum mass, while the infall time refers to when the subhalo first becomes part of the FOF group of the main halo's progenitor.
For the MWs, \SI{52}{\percent} of the total mass accretion can be attributed to mass bound to merging objects at some point in time, compared to \SI{55}{\percent} for M31. The remainder comes from unresolved mergers and smooth accretion.
Our MWs exhibit a sharp peak in merger-associated mass accretion at early times, followed by a steep decline after the third bin at an infall time of \SI{4}{Gyr}. In the difference plot on the right, this is reflected in the higher normalised mass accretion in the first three bins compared to our M31s. Conversely, at times later than \SI{4}{Gyr}, the M31s display more accretion from both major and minor mergers, indicating that our MWs generally have a quieter recent merger history than our M31s.
Quantitatively, when combining all infall events across our MWs and M31s, the median infall time of mass associated with major mergers (mass ratio $\mu > 0.25$) is at $t_{50,\mu>0.25}=\SI{3.3}{Gyr}$, which is $\SI{0.4}{Gyr}$ earlier than for the M31s ($t_{50,\mu>0.25} = \SI{3.7}{Gyr}$). Similarly, the median infall time for minor mergers ($0.1<\mu<0.25$) is $t_{50,0.1<\mu<0.25} = \SI{3.6}{Gyr}$ for the MWs, which is $\SI{0.7}{Gyr}$ earlier than for the M31 analogues ($t_{50,0.1<\mu<0.25} = \SI{4.3}{Gyr}$). The bootstrap errors, estimated by resampling the merger events, are less than \SI{0.16}{Gyr}, making the differences significant to $2.4\sigma$ for major mergers and $4.7\sigma$ for minor mergers.
We also examined the time of the last major merger (with mass ratio $\mu > 0.25$) of our MWs and M31s. The average time of the last major merger (LMM, $\mu > 0.25$) shows no significant difference between the two haloes: the MWs have an average $t_\text{LMM}$ of \SI{3.13}{Gyr} while the M31s have \SI{3.15}{Gyr}. However, $t_\text{LMM}$ is a rather stochastic quantity, with a large simulation-to-simulation scatter of \SI{2}{Gyr}, which could obscure any small signal. %

If we define GES-like mergers as satisfying:
\begin{enumerate}
\item the last merger with mass ratio $\mu > 0.2$;
\item the object was fully destroyed [2,6] Gyr after the Big Bang;
\item the mass ratio is $0.2<\mu<0.5$;
\end{enumerate}
we find that 72 out of our 224 simulations have such a merger.
Since the behaviour at late times is also interesting, we also searched for an LMC-like object, which we defined as
\begin{enumerate}
    \item  the LMC-subhalo is the most massive at the present day
    \item  it first crossed the MW's $r_\text{200c}$ more than \SI{0.5}{Gyr}~ago.
    \item the mass ratio is $0.1 < \mu < 0.2$.
    \item its peak mass is larger than \SI{5}{\percent} of the mass of the main progenitor at present day.
\end{enumerate}
In this case, we find that 28 out of our 224 simulations have such an LMC. Furthermore, 12 out of 224 simulations have both an LMC-like object and a GES-like merger.
\begin{figure}
    \centering
    \includegraphics[width=\linewidth]{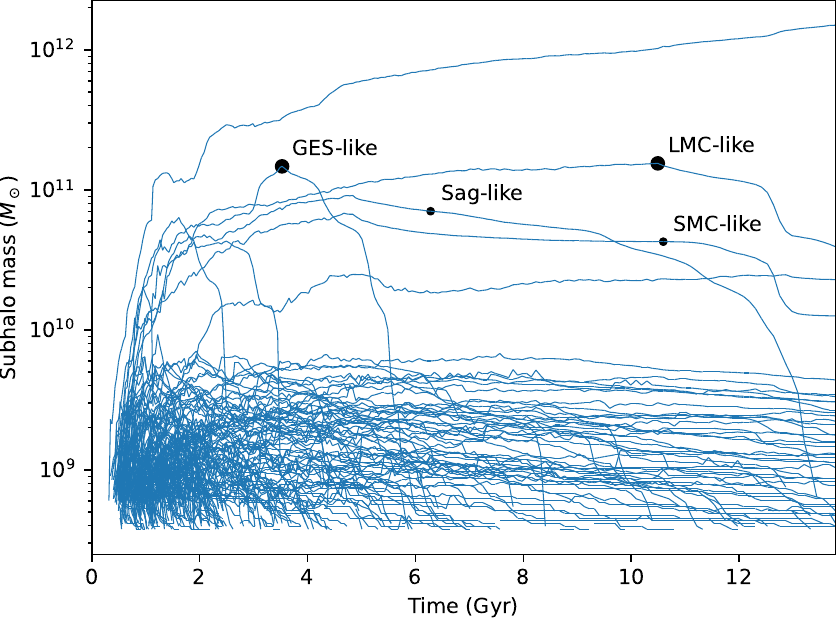}
    \caption{An example of the mass growth history of all the subhalos that end up in the Milky Way's FOF group. The top curve corresponds to the main progenitor of the Milky Way. This particular simulation (C10 S910 in the notation of \citetalias{wempeConstrainedCosmologicalSimulations2024}) has both a GES and an LMC analogue, which are indicated with the large black circles. A video of this particular simulation is available at \url{https://ewoudwempe.com/video_lmcsmclike.mp4}.}
    \label{fig:subhalosmassovertime}
\end{figure}

The mass growth history of one of these cases is shown in \cref{fig:subhalosmassovertime}. The LMC analogue in this simulation actually comes with a companion, an SMC-analogue, and there is also a merger resembling a Sagittarius analogue and an SMC. This simulation does not, however, feature any major or minor mergers for M31 in the last \SI{8}{Gyr}. The other 11 simulations with both an LMC-like and GES-like object have less quiescent merger histories for M31, with one experiencing a minor merger \SI{2}{Gyr} ago as proposed to have taken place for M31 by \citet{dsouzaAndromedaGalaxysMost2018}. However, the MWs do not host an LMC-SMC pair in these realisations.

We quantify the effect of having a GES-like merger or LMC-like merger on the growth times of the MW halo in \cref{tab:massgrowthtimes}.
A GES-like merger biases towards an earlier $a_{50}$ (MWs with a GES-like merger have $a_{50} = \num{0.43(7:9)}$, while without a GES requirement we find $a_{50} = \num{0.44(8:9)}$), which makes sense, given that our selection criteria require no major mergers (and thus no merger-associated growth) after the GES-like merger.
Requiring an LMC-like object does not significantly bias $a_{50}$: although one would expect to see the effect of the LMC in the growth history, in our definition, the LMC is required to be the most massive object at the present day.
Because of that, any late major mergers with mass ratio $\mu>0.2$ are removed from the sample, which biases one towards earlier assembly.
We find 12 simulations that have both a GES-like merger and an LMC analogue. Note also the reduced scatter in $a_{50}$ and $a_{10}$ when putting such constraints on the merger histories.

\begin{table}
\centering
\caption{Mean and 16/84th percentiles of mass growth histories.\label{tab:massgrowthtimes}}
\begingroup
\renewcommand{\arraystretch}{1.5}
\begin{tabular}{l r r r r}
\toprule
Selection & $N$ & $a_{10}$ & $a_{50}$ & $a_{90}$ \\
\midrule
Milky Way & 224 & \num{0.22(0.05:0.04)} & \num{0.44(0.08:0.09)} & \num{0.77(0.08:0.09)} \\
w. LMC & 28 & \num{0.22(0.05:0.05)} & \num{0.45(0.07:0.07)} & \num{0.76(0.08:0.08)} \\
w. GES & 72 & \num{0.21(0.04:0.03)} & \num{0.43(0.07:0.09)} & \num{0.80(0.07:0.08)} \\
w. LMC+GES & 12 & \num{0.19(0.03:0.03)} & \num{0.42(0.06:0.06)} & \num{0.75(0.08:0.10)} \\
M31 & 224 & \num{0.24(0.04:0.04)} & \num{0.48(0.09:0.10)} & \num{0.79(0.11:0.10)} \\
TNG-MW-mass & 52 & \num{0.26(0.04:0.04)} & \num{0.52(0.18:0.12)} & \num{0.78(0.14:0.13)} \\
TNG-M31-mass & 53 & \num{0.25(0.05:0.05)} & \num{0.51(0.11:0.11)} & \num{0.79(0.14:0.13)} \\
\bottomrule
\end{tabular}
\tablefoot{
The number $N$ of realisations (or objects) satisfying a certain selection criterion, and the scale factors at which \SIlist{10;50;90}{\percent} of their present-day $M_\text{200c}$'s were attained.
}
\endgroup
\end{table}

\subsection{Understanding the growth in an LG environment}

\begin{figure*}    \centering
    \includegraphics[width=0.99\linewidth]{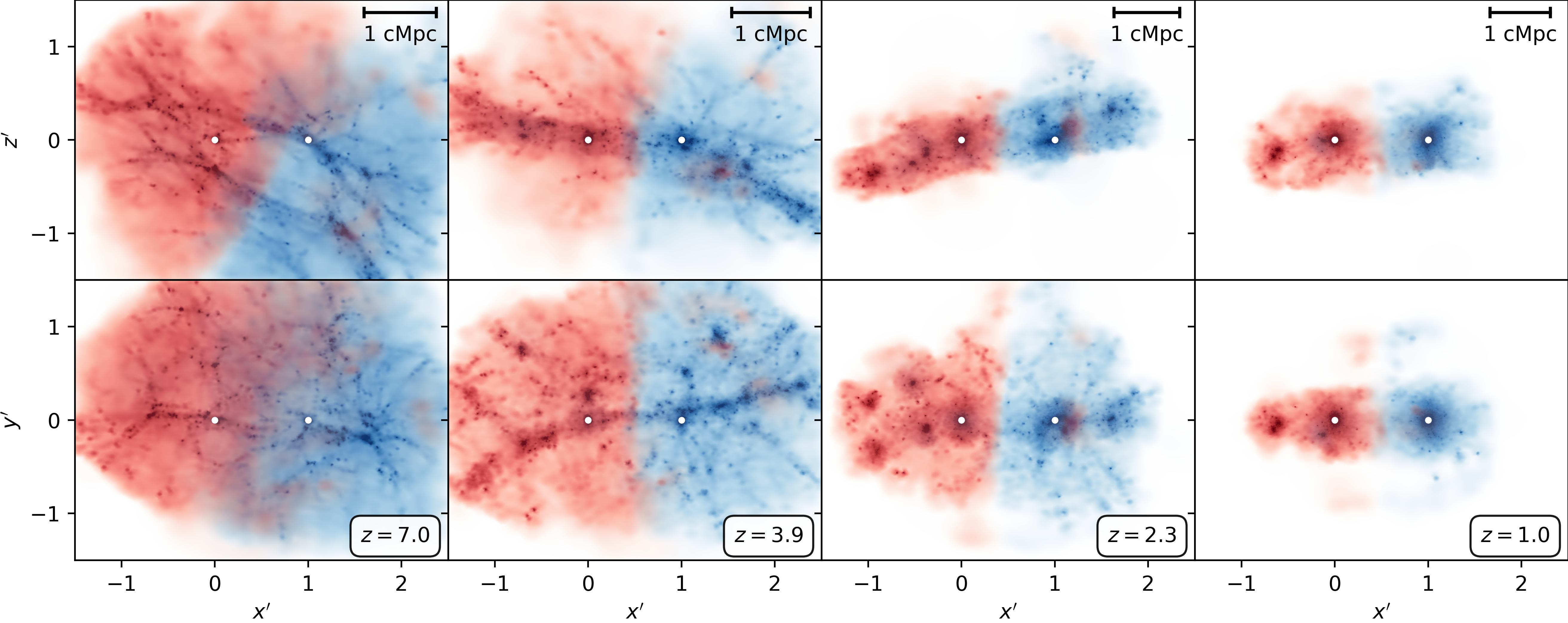}
    \caption{The particles that end up in the MW subhalo (in red) and in the M31 (in blue) at present day, traced back to various redshifts for one of our realizations. Each column shows a different redshift. The coordinate system has been rotated and scaled such that the MW and M31 main progenitors are at (0,0,0) and (1,0,0) respectively at all times (indicated with white dots). The top row shows an approximately edge-on view of the Local Sheet, while the bottom row shows a face-on view.
    \label{fig:accretion_tracing}
    }
    \vspace{0.5em}
    \includegraphics[width=0.99\linewidth]{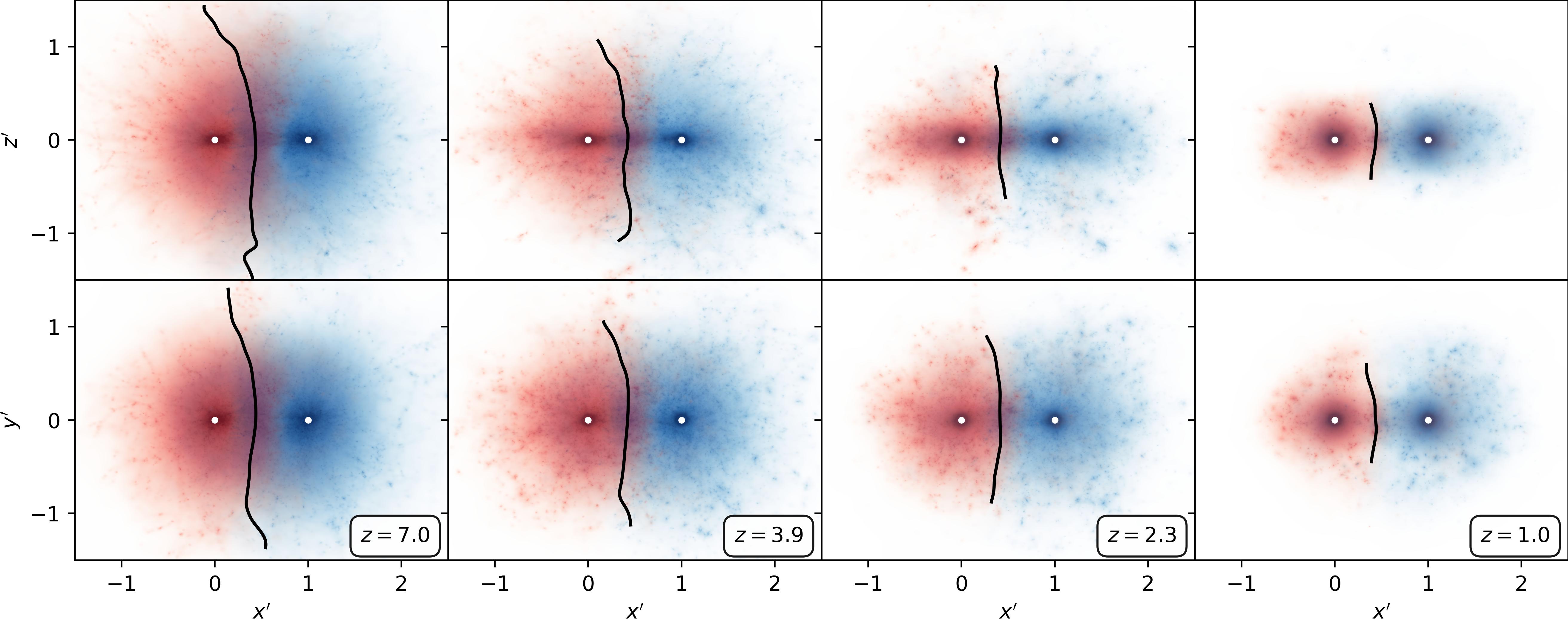}
    \caption{Like \cref{fig:accretion_tracing}, but now for the posterior mean of our simulations, a `stack' of all LG resimulations. The black lines are the isocontours at which a particle is equally likely to end up in the MW or M31. A video version of this figure is available at \url{https://ewoudwempe.com/progenitor_particles_stack.mp4}
    \label{fig:accretion_tracing_mean}
    }

\end{figure*}

To further investigate how the MW and M31 have grown in our LG simulations, in the top panels of \cref{fig:accretion_tracing} we consider the simulation shown in \cref{fig:subhalosmassovertime}, and identify the positions at earlier times of the particles that end up in the MW and M31 FoF groups at the present day. In 65 out of 224 cases the MW and M31 halos end up in the same FoF group, and in these cases, the satellites are assigned to the closest halo. %
In this figure, we have rotated and scaled the coordinate system such that the MW's main progenitor is positioned at $(x', y', z') = (0,0,0)$, and M31's main progenitor is at $(1,0,0)$ at all times.
This leaves one free angle (the rotation around the MW-M31 axis), which we set such that the $z'$-axis aligns as closely as possible with the north pole of the Local Sheet (at supergalactic coordinates $(L,B) = (\SI{241.74}{\degree}, \SI{82.05}{\degree})$, \citealt{mccallCouncilGiants2014}), which roughly aligns with the Supergalactic north pole (the separation is \SI{7.95}{\degree}).
As a result, the $z'$-axis maintains close alignment with the Local Sheet north pole. This is not exact because of M31's slight elevation above the Local Sheet, but the misalignment is minor, averaging \SI{23}{\degree} at $z=7$, and reducing to \SI{12}{\degree} at $z=0$.
The left column shows the distribution at redshift $z\sim 7$, which reveals a few narrow filaments already forming with low $z'$ and a generally isotropic distribution around the two halos.
Then, at $z\sim 3.9$, a large part of the MW's material has collapsed into a sheet-like distribution , aligned with the $z'$ direction, with a filament forming along the MW-M31 axis. At $z=2$, the collapse along the $z'$ direction is mostly complete, and further growth occurs mostly from within the sheet. By $z=1$, this halo has accreted more than half of its material, and the distribution is again more axisymmetric around the MW-M31 axis.

Although this is a single simulation, this behaviour is similar across all LG realisations, as depicted in
\cref{fig:accretion_tracing_mean}. In this figure we stacked all the MW and M31 progenitor particles of all simulations to get a map of the posterior mean behaviour of the positions of the progenitor particles over time (note that all simulations are samples from the Local Group posterior of \citetalias{wempeConstrainedCosmologicalSimulations2024}, and hence stacking them gives the posterior mean field).
At $z=7$, the progenitor particle distribution is, on average, elongated along the $z'$-axis (which is roughly aligned with the supergalactic $z$-axis). Then up until $z\sim2$, it collapses along the $z'$-axis, becoming most compressed along this axis. After $z=2$, the halo material starts to virialise and the distribution becomes roughly spherically symmetric by $z=1$.
The separatrix between the MW and M31 material is shown with black lines. It lies at $x'=0.4$ along the MW-M31 axis at $z=7.0$; particles with $x' < 0.4$, are more likely to become part of the MW, and those with $x' > 0.4$ are more likely to become part of M31.
Its location at $x'=0.4$ stays roughly constant throughout cosmic time, with some scatter ranging from 0.37 to 0.45. Interestingly, this range may be expected from a simple computation of the location of the Lagrange point L1 in a system of two point masses (without rotation, as the tangential velocities are negligible) and with a mass ratio of 0.57 (which is the median value for our simulations).

Using \cref{fig:accretion_tracing_mean} we can point out the effect that suppresses the Milky Way's mass growth at late times: due to the higher mass of M31, the particles that are right of the separatrix will go to M31 instead of the Milky Way, and this especially suppresses late-time growth, when particles from larger initial distances dominate the mass growth. This is also seen in the unnormalised mass growth histories, shown in \cref{app:unnormalised}, which at early times largely coincide, and start to deviate at later times.

\section{Discussion and conclusions}

We have found that MW halos in our constrained LG simulations generally form earlier and grow more slowly at late times than their isolated counterparts. This is principally due to the presence of M31 (which is more massive by a factor of approximately 1.75 in the median) but may also be due in part to the quiet surrounding Hubble flow. We also find the growth of both galaxy halos to be quite anisotropic in our constrained simulations. Protohalo material first collapses into a planar distribution, which is roughly aligned with the Local Sheet, and then, after $z\sim2$, starts to virialise, ending up in a significantly less flattened configuration already by $z\sim1$.

Quantitatively, our MWs reach half their final mass at ${a_\text{50,MW} = \num{0.44(8:9)}}$, while for M31 this is ${a_\text{50,M31}= \num{0.48(9:10)}}$, while MW-mass and M31-mass isolated halos from TNG50-DMO have $(a_\text{50,MW-mass},a_\text{50,M31-mass}) = (\num{0.52(18:12)}, \num{0.51(11:11)})$.
Requiring the last major merger to be GES-like biases the Milky Way to ${a_\text{50,MW | GES}=\num{0.43(7:9)}}$.
The early and late time behaviour can be quantified by $(a_\text{10,MW}, a_\text{10,M31}) = (\num{0.22(5:4)}, \num{0.24(4:4)})$ and $(a_\text{90,MW}, a_\text{90,M31})$ $= (\num{0.77(8:9)}, \num{0.79(11:11)})$.

These values are not very different from those of \citet{buchMilkyWayestCosmological2024} for MW-like isolated halos. They find $a_{50} = 0.49$ for a similar sample of MW-mass objects without the requirement of an LMC (in which case $a_{50} = 0.57$), although in the mean, our MWs reach their half-mass slightly earlier. In the ELVIS suite of LG analogues, \citet{garrison-kimmelStarFormationHistories2019} and subsequently \citet{santistevanFormationTimesBuilding2020} find that for isolated halos $a_{10} \sim 0.25$,  while for LG-like hosts they find $a_{10} \sim 0.2$, in agreement with our own results. These authors argue that at later times,
$a \gtrsim 0.3$, the mass growth histories of isolated and LG-like hosts are more similar, which might indicate a role of the initial overdensity associated with LG in comparison to a random MW or M31 progenitor.

Similar findings are reported for the HESTIA simulations of LG-like pairs inside constrained simulations of the large-scale structure \citep[with constraints on scales larger than $\SI{\sim4}{Mpc}$][]{libeskindHESTIAProjectSimulations2020}. In these simulations, on average $a_{50} = 0.42$, indicating that these halos form roughly one billion years earlier than in unconstrained similarly massed MW/M31 analogues (whose $a_{50} \sim 0.48$).
Note however that these authors make no distinction between MW and M31 halos in their estimates of $a_{50}$, so these numbers reflect the effect of a group environment and not differences between the two systems.
In a sample of LGs extracted from CLUES \citep{carlesiConstrainedLocalUniversE2016a},  \citet{carlesiMassAssemblyHistory2020} found that the median half-mass time of halos of galaxies in LGs in constrained realizations is also earlier than in LGs without constraints, although the difference is small, and of the order of 0.5 Gyr both for the MWs and for the M31s.
\citet{carlesiMassAssemblyHistory2020} notice that their MWs reach their half mass earlier than their M31s (by 0.25 Gyr), but attribute this difference to mass.
We, on the other hand, find that our MWs have also a smaller-than-average half-mass time than isolated MW-mass galaxies.
The difference between MW and M31 we find is also larger, $\expval{t_\text{50,M31}} - \expval{t_\text{50,MW}} = \SI{0.6}{Gyr}$.
This could plausibly be explained by the separation between our LG pairs, which is \SI{806(124:133)}{kpc}, whereas \citet{carlesiMassAssemblyHistory2020} allows for a much larger range of \SIrange{0.5}{1.9}{Mpc}, which may imply a larger separation between MW-M31 on average in their sample (see e.g. Table 7 of \citealt{carlesiConstrainedLocalUniversE2016a}), that in turn reduces the effect of M31 on the MW.

These findings may have implications for the formation of the Galactic disk, as explored by \citet{semenovHowEarlyCould2024} using the TNG50 simulations. It appears to be relatively rare (less than 10\%) for disk galaxies of MW mass in typical environments to have assembled  a significant fraction of their mass within the first few Gyr and to have formed a disk that survives until the present day. Nevertheless, this seems to have happened in the real MW.
However, we find that once the LG structure and its local environment environment are correctly reproduced, such behaviour is quite common.
Increasing the chance of such an early assembly history requires i) a quiet environment surrounding the LG, and ii) a relatively large mass ratio of M31 to the MW; both factors reduce the likelihood of significant late-time growth. %

Possible improvements to the analysis presented here would include using a larger sample to increase the numerical precision of our estimates. This could also allow us to be stricter in selecting close analogues of MW and M31, for example, by identifying LMC-like objects that also reproduce the LMC's present-day distance and approximate orbit, although for a high quality match, a preferable approach would be to add the LMC (or M33) as an extra constraint to the \citetalias{wempeConstrainedCosmologicalSimulations2024} methodology. Another possibility would be to constrain the spin orientation of the inner MW and M31 haloes to match the observed orientations of the two galaxies. A further obvious improvement would be to carry out hydrodynamical simulations, as these would allow us to follow the mass growth of all the different constituents of the galaxies, i.e. cold gas, hot gas and stars, as well as the dark matter. We are currently working on such a suite of simulations, and hope to report on these in future work.

\begin{acknowledgements}
This work has been financially supported by a Spinoza Prize from NWO to AH.
EW thanks Dylan Nelson for providing access to the TNG simulations, Anna Genina and Victor Forouhar Moreno for helpful discussions, and Akshara for her moral support.
The Gadget simulations were performed on the Freya cluster at the Max Planck Computing and Data Facility. The constrained simulations were enabled by resources provided by the Swedish National Infrastructure for Computing (SNIC) at the PDC Center for High Performance Computing, KTH Royal Institute of Technology, partially funded by the Swedish Research Council through grant agreement no. 2018-05973.
JJ and GL acknowledge support from the Simons Foundation through the Simons Collaboration on "Learning the Universe". This work was made possible by the research project grant "Understanding the Dynamic Universe," funded by the Knut and Alice Wallenberg Foundation (Dnr KAW 2018.0067). Additionally, JJ acknowledges financial support from the Swedish Research Council (VR) through the project "Deciphering the Dynamics of Cosmic Structure" (2020-05143).
GL acknowledges support from the CNRS-IEA "Manticore" project. This work is done within the Aquila Consortium\footnote{\url{https://www.aquila-consortium.org/}}.\\
\textit{Software:}
During this work, we have made use of various software packages. This includes \borg  \citep{jascheBayesianPhysicalReconstruction2013,jaschePhysicalBayesianModelling2019}, \textsc{Gadget-4} \citep{springelSimulatingCosmicStructure2021}, \textsc{jax} \citep{jax2018github}, \textsc{numpyro} \citep{phanComposableEffectsFlexible2019}, \textsc{xarray} \citep{hoyer2017xarray,hoyer_2023_10023467}, \textsc{numpy} \citep{harris2020array}, \textsc{scipy} \citep{2020SciPy-NMeth}, \textsc{matplotlib} \citep{Hunter:2007}, \textsc{h5py}\footnote{\url{https://www.h5py.org/}}, \textsc{astropy} \citep{astropycollaborationAstropyProjectBuilding2018,astropycollaborationAstropyProjectSustaining2022}, \textsc{jug} \citep{coelhoJugSoftwareParallel2017}.
\end{acknowledgements}

\bibliographystyle{aa}
\bibliography{references_zot,phd}

\appendix

\section{Upsampling of initial condition fields}
\label{app:icupsampling}
In \citetalias{wempeConstrainedCosmologicalSimulations2024}, we utilised a zoom scheme that involved splitting the initial condition field into two parts, a field for the low-resolution large box, $\vb*s_\text{LR}$ and a field for the high-resolution zoom region, $\vb*s_\text{HR}$.
The zoom region has a box size of \SI{10}{Mpc}, with 64 elements on a side, giving $\vb*s_\text{HR}$ a mesh cell size of $\frac{\SI{10}{Mpc}}{64} = \SI{156.25}{kpc}$.
For our resimulations, we use a grid twice as fine for the zoom region while leaving the outer region unchanged. Therefore, we upsample the $\vb*{s}_\text{HR}$ field, from its original $64^3$ size to a $128^3$ grid, preserving all the Fourier modes of the original $\vb*{s}_\text{HR}$ field, and randomly generating the modes at frequencies larger than the Nyquist frequency (per axis) of $\vb*{s}_\text{HR}$. %

To accomplish this, we combine two fields: our original field $\vb*{s}_\text{HR}$, which sets the larger scale modes, and a randomly generated field ($\vb*s_\text{ss}\sim \mathcal{N}(0,1)$) that sets the smallest scale Fourier modes.
To ensure that the final field is real-valued, we use the Hartley transform which guarantees a real-valued output for a real-valued input. The Hartley transform $\vu*{s}$ of a real field $\vb*{s}$ is computed from the Fourier transform ($\vu*s = \mathcal{H} \vb*{s} = \Re{\mathcal{F}\vb*s} - \Im{\mathcal{F}\vb*s}$).
Our new field's modes are set by taking $\vu*{s}_\text{HR}$ modes at frequencies less than the Nyquist frequency of $\vb*s_\text{HR}$ (the large scales) and from $\vu*{s}_\text{ss}$ at higher frequencies (the small scales).
At the Nyquist frequencies, the modes are split equally across the positive and negative Fourier modes in the new field, to ensure that, if $\vb*s_\text{ss}=0$, the process corresponds to pure trigonometric interpolation.
Hence, when one, two or three indices are $\pm i_\text{Ny}$, the contribution of $\vu*s_\text{HR}$ is weighted by $a = \frac{1}{2}$, $\frac{1}{4}$, and $\frac{1}{8}$, respectively.
Since the variance of a mode $X = a \mathcal{N}(0,1) + b \mathcal{N}(0,1)$ is given by $\sigma^2 = a^2+b^2$, to conserve unit variance at these modes, the coefficient $b$ for the small-scale field $\vu*{s}_\text{ss}$ is set to $b = \sqrt{1-a^2}$, resulting in $\sqrt{\frac{3}{4}}$, $\sqrt{\frac{15}{16}}$ and $\sqrt{\frac{63}{64}}$ for the three cases respectively.
Summarising, the modes of the new $\vb*{s}_\text{HR}^\text{new}$ field are:
\begin{align*}
    \vu*{s}_\text{HR}^\text{new} = \begin{cases}
        \vu*{s}_\text{HR} &\hspace{-0.5em}\parbox{16em}{if the absolute values of all indices are less than $i_\text{Ny}$,}\\[0.8em]
        \vu*{s}_\text{ss} &\hspace{-0.5em}\parbox{16em}{if the absolute value of any index exceeds $i_\text{Ny}$,}\\[0.0em]
        \frac{1}{2}\vu*{s}_\text{HR} + \sqrt{\frac{3}{4}}\vu*{s}_\text{ss} &\hspace{-0.5em}\parbox{16em}{if exactly one index is $\pm i_\text{Ny}$,}\\[0em]
        \frac{1}{4}\vu*{s}_\text{HR} + \sqrt{\frac{15}{16}}\vu*{s}_\text{ss} &\hspace{-0.5em}\parbox{16em}{if exactly two indices are $\pm i_\text{Ny}$,} \\[0em]
        \frac{1}{8}\vu*{s}_\text{HR} + \sqrt{\frac{63}{64}}\vu*{s}_\text{ss} &\hspace{-0.5em}\parbox{16em}{if all indices are $\pm i_\text{Ny}$,}
    \end{cases}
\end{align*}
where the indices range from $-N_\text{ss}/2$ to $N_\text{ss}/2$, $i_\text{Ny} = N_\text{HR}/2$ is the Nyquist index of the $\vu*{s}_\text{HR}$ field, and the last three cases apply only if no index exceeds $i_\text{Ny}$ in absolute value.

\section{Autocorrelation analysis}
\label{app:autocorrelations}

In this work, we use initial conditions inferred from Hamiltonian Monte Carlo (HMC) chains with relatively long autocorrelation lengths. As a result, some quantities extracted from the simulations may exhibit autocorrelation, which increases the numerical error in the posterior means reported in this paper. For independent samples of a quantity $Q$, the standard deviation of the mean is $\sigma_{\expval{Q}} = \frac{\sigma_Q}{\sqrt{N}}$. For correlated samples, this becomes $\sigma_{\expval{Q}} = \frac{\sigma_Q}{\sqrt{N_\text{eff}}}$, where $N_\text{eff}$ is the effective sample size. The effective sample size is calculated as $N_\text{eff} = N / (1 + 2\sum_{d=1}^\infty \rho(d))$, where $\rho(d)$ is the autocorrelation function of $Q$ at lag $d$.

In \citetalias{wempeConstrainedCosmologicalSimulations2024}, the autocorrelation function of the initial condition fields, $\rho_{\vb*{s}}(d)$, was found to follow an exponential decay, $\rho(d) = e^{-d/L}$, with autocorrelation length $L$. Fitting the model to the 12 HMC chains resulted in an effective sample size (ESS) of $N_\text{eff}=20$ for the initial condition field.\footnote{This differs from the $N_\text{eff} = 27$ reported in \citetalias{wempeConstrainedCosmologicalSimulations2024} due to a programming error in the effective sample size calculation in that paper. The results of \citetalias{wempeConstrainedCosmologicalSimulations2024} are unaffected, as all quoted uncertainties reflect the posterior widths, which are independent of the effective sample size. The corrected effective sample size of $N_\text{eff} = 20$ inflates the predictive uncertainties on inferred quantities by \SI{2.5}{\percent}, compared to \SI{1.7}{\percent} with $N_\text{eff} = 27$. This difference remains negligible.} For this paper, we have run the same 12 chains longer, producing an effective sample size of the initial condition field of $N_\text{eff} = 32$. From these chains, we extract 224 realisations.

It is important to note that for quantities sensitive to the behaviour on small scales (e.g. below 1 Mpc), the correlation lengths between samples will be smaller, hence the 224 realisations could be seen in this context as (quasi)independent. We explain below how to quantify the numerical uncertainties on the various quantities discussed in this paper taking into account the quasi-independence of our samples.

As an example, we determine the effective sample size of $a_{50}$. We assume an exponential autocorrelation function $Ae^{-d/L_s}$ and that autocorrelation lengths $L_s$ on small scales are a factor $f$ smaller than those for the initial conditions, i.e $L = f \times L_s$. To estimate $L_s$, we fit the autocorrelation function using NumPyro. We assume $a_{50}$ follows a log-normal distribution, $\ln{a_{50}} \sim \mathcal{N}(\mu, \sigma^2)$, with an exponentially declining autocorrelation function:
\begin{equation}
\rho_{a_{50}}(d) = \expval{(a_{50}[i] - \mu)(a_{50}[i+d] - \mu)} = A e^{-d/L_s}.
\end{equation}
We find that $f=\num{0.70(17)}$ and $A=\num{0.41(7)}$, giving an effective sample size of $N_\text{eff}=\num{73(13)}$. The Monte Carlo standard error on $a_{50}$ is 0.01 for both the Milky Way and M31, and the earlier average assembly reported in \cref{sec:massgrowthhistory} is significant at the \SI{1.7}{\percent} level.

For the $a_{50}$ of the mass bound to the main subhalo, the difference between MW and M31 is even more significant, at \SI{0.4}{\percent}. In this case, we removed the 65 cases in which the MW and M31 ended up in the same FoF group, to avoid potential biases due to the differing behaviour in the Subfind unbinding procedure in the same-FoF-group and different-FoF-group scenarios. Specifically, the Subfind unbinding procedure only considers particles within the same FoF group, meaning the tidal effect of M31 lowers the MW mass in the same-FoF-group case, but not in the different-FoF-group scenario. %

\section{Unnormalised mass growth histories}
\label{app:unnormalised}
\begin{figure}[htpb]
    \centering
    \includegraphics[width=\linewidth]{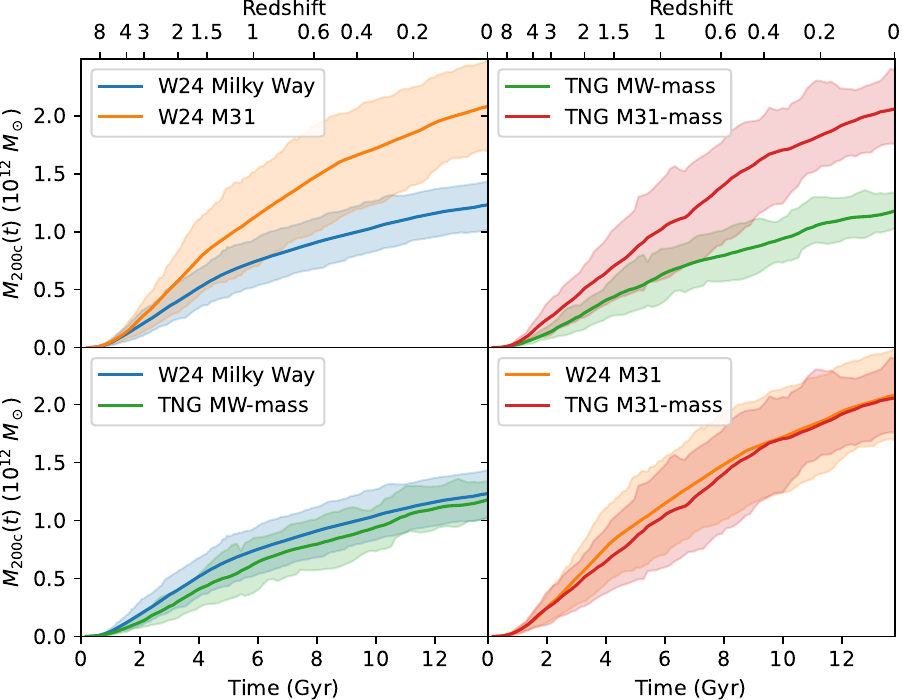}
    \caption{Like \cref{fig:mencevolution}, but unnormalised. Note that at early times, the MW and M31 had a very similar mass growth history, diverging from $z\sim4$. The larger scatter in the present-day M31 mass reflects the larger (observational) uncertainties on its mass from \citetalias{wempeConstrainedCosmologicalSimulations2024}. The top-right panel shows the same, but for MW-mass and M31-mass galaxies from an unconstrained simulation, and the bottom panels show comparisons of the MWs and M31s from the constrained simulations to isolated MW/M31-mass haloes.}
    \label{fig:unnormalised}
\label{LastPage}
\end{figure}

Instead of the normalised growth histories shown in \cref{fig:mencevolution}, one can also consider the unnormalised growth histories, which are shown in \cref{fig:unnormalised}. Note that the Milky Way and M31 have very similar mass assembly histories at early times,  diverging only around redshift 4.%

\end{document}